\documentclass[aps,prx,twocolumn,showpacs,superscriptaddress,longbibliography]{revtex4-2}
\usepackage{amsmath,amssymb,graphicx}
\usepackage{graphicx,epstopdf}
\usepackage{gensymb}
\usepackage[dvipsnames]{xcolor}
\usepackage[T1]{fontenc}
\usepackage[utf8]{inputenc}
\usepackage{xspace}
\usepackage{amsfonts}\usepackage{amsthm}
\usepackage{mathrsfs}\usepackage[title]{appendix}
\usepackage{textcomp}
\usepackage{booktabs}
\usepackage{algpseudocode}\usepackage{listings}
\usepackage{multibib}
\usepackage{color}
\usepackage[colorlinks,bookmarks=false,citecolor=darkblue,linkcolor=red,urlcolor=blue]{hyperref}

\definecolor{darkred}{rgb}{0.7,0.0,0.0}

\definecolor{darkblue}{rgb}{0,0.02,0.45}

\definecolor{darkgreen}{rgb}{0.02,0.45,0.0}

\definecolor{violet}{rgb}{0.8,0.2,0.6}

\newcommand{\LCS}{LaCrSb$_{3}$\xspace}
\newcommand{\HIIa}{$\textbf{H} \parallel {\bf a}$\xspace}
\newcommand{\HIIb}{$\textbf{H} \parallel {\bf b}$\xspace}
\newcommand{\HIIc}{$\textbf{H} \parallel {\bf c}$\xspace}

\definecolor{viol}{rgb}{0.7, 0.4, 1}
\usepackage{placeins}
\usepackage[normalem]{ulem}
\usepackage{soul}

\usepackage{hyperref}

\definecolor{mygray}{cmyk}{0, 0, 0, 0.3}
\newcommand{\out}[1]{{\color{mygray}\sout{#1}}}

\usepackage{float}

\begin{document}

\title{Signatures of unconventional magnetism in the layered metallic ferromagnet LaCrSb$_3$ from ferromagnetic resonance spectroscopy}

\author{J.~J.~Abraham}
\affiliation{Leibniz Institute for Solid State and Materials Research Dresden 
	Helmholtzstr. 20, D-01069 Dresden, Germany}
\affiliation{Institute for Solid State and Materials Physics, TU Dresden, D-01062 Dresden, Germany}
\author{S. Samanta}
\affiliation{School of Physics, Indian Institute of Science Education and Research, Thiruvananthapuram 695551, India}
\author{R. Kolay}
\affiliation{School of Physics, Indian Institute of Science Education and Research, Thiruvananthapuram 695551, India}
\author{V. Singh}
\affiliation{School of Physics, Indian Institute of Science Education and Research, Thiruvananthapuram 695551, India}
\author{R. Nath}
\email{rameshchandra.nath@gmail.com}
\affiliation{School of Physics, Indian Institute of Science Education and Research, Thiruvananthapuram 695551, India}
\author{B.~B\"uchner}
\affiliation{Leibniz Institute for Solid State and Materials Research Dresden 
	Helmholtzstr. 20, D-01069 Dresden, Germany}
\affiliation{Institute for Solid State and Materials Physics, TU Dresden, D-01062 Dresden, Germany}
\affiliation{Institute for Solid State and Materials Physics and W{\"u}rzburg-Dresden Cluster of Excellence ctd.qmat, TU Dresden, D-01062 Dresden, Germany}
\author{V.~Kataev}
\email{v.kataev@ifw-dresden.de}
\affiliation{Leibniz Institute for Solid State and Materials Research Dresden 
	Helmholtzstr. 20, D-01069 Dresden, Germany}
\author{A.~Alfonsov}
\email{a.alfonsov@ifw-dresden.de}
\affiliation{Leibniz Institute for Solid State and Materials Research Dresden 
	Helmholtzstr. 20, D-01069 Dresden, Germany}

\date{\today}

\begin{abstract}
\LCS is a metallic ferromagnet with a layered crystal structure demonstrating intriguing electronic and magnetic properties, such as large anomalous Hall effect, strong canting of the spin lattice, and a peculiar spin-reorientation transition. Here, we report the results of the temperature-dependent x-ray diffraction, static magnetization, and in particular electron spin resonance (ESR) and ferromagnetic resonance (FMR) experiments carried out over a wide range of frequencies, magnetic fields, and temperatures. Though x-ray data reveals no structural transition down to 15\,K, a strong magneto-elastic coupling is detected across the ferromagnetic transition at $T_{\rm C} \simeq 126$\,K. ESR results indicate a presence of the quasi-static short-range correlations extending far above $T_{\rm C}$, which is a typical fingerprint of the low-dimensional magnetism. The frequency-field diagram of the FMR modes mapped below $T_{\rm C}$ strongly suggests presence of two magnetic sublattices in \LCS . A quantitative understanding of the FMR excitations was achieved within a phenomenological model of interacting orthogonal ferro- and antiferromagnetic sublattices which was earlier proposed to explain unusually strong spin canting observed by neutron diffraction [\href{https://doi.org/10.1103/PhysRevLett.89.107204}{E.~Granado {\it et al.}, Phys. Rev.Lett.\,{\bf 89}, 107204 (2002)}]. The FMR results corroborate this scenario and call for the development of the underlying microscopic model of unconventional magnetism in \LCS.      
\end{abstract}

\maketitle

\section{Introduction} 
\begin{figure} 
	\includegraphics[width=0.8\columnwidth]{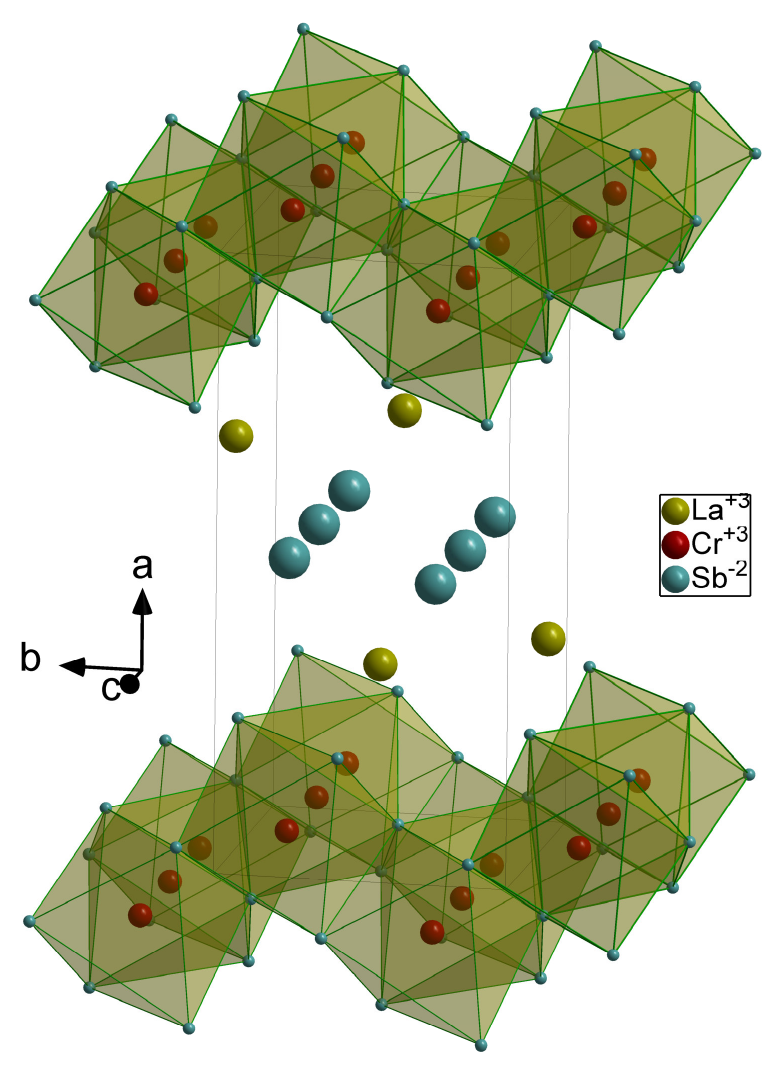}
	\caption{Orthorhombic crystal structure of \LCS, space group $Pbcm$ (No.\,57). Cr cations are octahedrally coordinated by Sb ligands. Octahedra share edges along the $b$-axis and faces along the $c$-axis forming a 2D network in the $bc$-plane. The Cr-Sb layers are well separated along the $a$-axis by Sb and La sheets.  }
	\label{figure:structure}
\end{figure}

The ongoing research on two-dimensional (2D) layered compounds has drawn significant attention due to their intriguing physical properties and potential for a wide variety of applications. This very rich class of materials spans layered superconductors, such as celebrated high-$T_{\rm c}$ cuprates, iron pnictides, transition metal (TM) dichalcogenides, organic layered superconductors, etc. \cite{Klemm2011}, van der Waals compounds \cite{Novoselov2004,Geim419,Ajayan2016,Novoselov9439,Gibertini2019,Yang2021,Xu2022,Liu2023}, kagome materials \cite{Yin2022,Wang2023}, and many others.   
The discovery of 2D TM-based ferromagnets such as, e.g., CrGeTe$_3$, CrI$_3$, Fe$_3$GeTe$_2$, MnSe$_2$, and VSe$_2$, has provided a new platform for fundamental studies of magnetism in reduced spatial dimensions as well as has opened up exciting opportunities for developing next-generation spin-based devices~\cite{Gong265,Bonilla289,Geim419,Novoselov9439}.
A few recent examples of the metallic 2D TM-based compounds hosting interesting physical properties are  CrI$_3$ and AgCrSe$_2$ demonstrating large magnetoresistance \cite{Wang2516,Takahashi054602}, Mn$_{1.4}$Fe$_{3.6}$Si$_3$ with its  large negative thermal expansion coefficient \cite{Singh033902}, as well as the metallic rare-earth antimonides $R$CrSb$_3$ ($R$ = La, Ce, Pr, Nd) featuring anomalous Hall effect and the magnetoresistance \cite{Leonard265,Crerar2780,Inamdar132410,Inamdar2766,Kumar2021,Paul094427}.  Recently, the $^{139}$La nuclear magnetic resonance on an itinerant ferromagnet LaCrGe$_3$ revealed a ferromagnetic quantum critical point under pressure~\cite{Rana214417,Rana174426}. LaCrGe$_3$ also shows a large anomalous Hall conductivity when a magnetic field is applied along the magnetic easy axis~\cite{Sariket2025}. Its sister compound, \LCS (Fig.~\ref{figure:structure}), is a metallic ferromagnet as well, with the reported Curie ordering temperature $T_{\rm C} \simeq 125$\,K \cite{Raju3630}. It features anisotropic magnetic and transport properties~\cite{Jackson014421,MacFarlane374,Yang024419}, and in particular giant anomalous Hall conductivity \cite{Kumar2021}. The magnetic order is of the easy-plane type with a small in-plane easy-axis anisotropy which changes its direction from the $c$- to the $b$-crystal axis upon decreasing the temperature below the spin-reorientation transition $T^\ast \simeq 95$\,K \cite{Jackson014421,Granado107204}.  The critical analysis of magnetization revealed that the system falls between 3D Heisenberg and mean-field models \cite{Yang024419}. Density functional theory calculations put forward a kinetic energy-driven double exchange mechanism as the main reason for a relatively high value of $T_{\rm C}$ despite the low-D character of the spin system in \LCS \cite{Paul094427}.

One puzzling, unconventional aspect of magnetism of \LCS revealed by magnetic neutron diffraction is a significant canting of the ferromagnetically (FM) ordered spin structure with the angle between neighboring Cr spins amounting to 36$^\circ$ \cite{Granado107204}. It was argued that such a large canting is unlikely to be caused by the spin-orbit effects. Instead, it was conjectured to be due to a presence in \LCS of interacting FM and antiferromagnetic (AFM) sublattices originating from coexisting localized and itinerant spins \cite{Granado107204}.
In order to verify that we have performed measurements of electron spin resonance (ESR) and ferromagnetic resonance (FMR) on poly- and single-crystalline samples of \LCS in the broad ranges of excitation frequencies, magnetic fields and temperatures. The studied samples were thoroughly characterized by temperature-dependent x-ray diffraction and static magnetometry confirming their high quality. The x-ray data reveals a strong magneto-elastic coupling near $T_{\rm C}$ with a negative thermal expansion coefficient. ESR results give strong indication of persistent quasi-static FM correlations at temperatures far above $T_{\rm C}$ typical for low-D spin systems. Below $T_{\rm C}$, the frequency-field diagram of the FMR modes (resonance branches) was mapped. At $T^\ast < T < T_{\rm C}$, only one FMR branch for a given field direction is present, as expected for a one sublattice ferromagnet, such as \LCS supposed to be. However, this branch splits below $T^\ast$ in two branches, requiring for explanation an assumption of a presence of at least two magnetic sublattices in this compound. To explain this observation, we have developed a phenomenological model of interacting orthogonal FM and AFM sublattices, as proposed in Ref.~\cite{Granado107204}, and solved numerically its spin-wave excitation spectrum in the frame of the linear spin wave theory (LSWT). For a certain set of model parameters, we could achieve a good quantitative agreement with experimentally observed FMR branches. Thus, our findings give strong support to the interpretation of the magnetic neutron diffraction and highlight a possible unconventional nature of magnetism in \LCS.              

\section{Experimental Details}
The single crystals of LaCrSb$_3$ were grown using the Sb flux, as reported previously~\cite{Jackson014421}. A mixture of elements in stoichiometric ratio La:Cr:Sb = 4:3:13 was heated in an evacuated silica tube at $1150^{\circ}\mathrm{C}$ for 24~hours and then slowly cooled to $750^{\circ}\mathrm{C}$ in a cooling rate of $3~^{\circ}\mathrm{C}$/hr. 
The as-grown crystals appeared as thin, plate-like structures with a characteristic silvery metallic luster. They are found to be stable in air for a long period of time. We crushed the single crystals and checked the phase purity of the powder sample using a PANalytical powder x-ray diffractometer with Cu$K_\alpha$ radiation ($\lambda_{\rm avg} = 1.5406~\text{\AA}$). To analyze the structural changes across the magnetic transition, temperature-dependent powder x-ray diffraction (XRD) measurements were performed in a temperature range of 15 to 300~K. For this purpose, an Oxford-PheniX closed-cycle helium cryostat was used as an attachment to the diffractometer. Rietveld refinement of the powder XRD data was performed using the FullProf software package~\cite{Carvajal55}.

The dc magnetization ($M$) as a function of temperature (1.8~K~$\leq T \leq 380$~K) and magnetic field (0~T~$\leq \mu_0 H \leq 7$~T) applied along different crystallographic directions was measured on single crystals using a superconducting quantum interference device (SQUID) (MPMS-3, Quantum Design) magnetometer.

ESR measurements at a fixed microwave frequency $\nu = 9.56$\,GHz were performed using a Bruker X-band spectrometer (EMX series). Magnetic fields were swept up to 0.9\,T, and a continuous-flow helium cryostat (Oxford Instruments) allowed precise control of the sample's temperature across a wide range from 4 to 300\,K. The spectrometer was equipped with a goniometer to achieve a proper alignment of the sample, with magnetic field parallel to its $a$-, $b$- or $c$-axes. Frequency swept ESR experiments in the range 1--67\,GHz. were carried out with a home-made ESR setup based on  a vector network analyzer (PNA-X from Keysight Technologies) continuously operational in this frequency range. Magnetic fields were generated using a superconducting optical magnet system from Oxford instruments (Spectromag). A coplanar waveguide with the sample attached to it was introduced into a $^4$He variable-temperature insert (VTI) integrated into the magnet system. The VTI enabled accurate control of the sample's temperature in the range 4.2--300\,K. Both studies were performed on a single crystalline sample, whose crystallographic axes were identified with the help of magnetometry measurements (see Appendix~\ref{section:additional_data}, Fig.~\ref{fig:MH}).

High-frequency/high-field ESR (HF-ESR) measurements were performed on a powder sample using a home-made HF-ESR spectrometer capable of operating over a broad range of frequencies and magnetic fields. The setup consists of a vector network analyzer (PNA-X from Keysight Technologies) equipped with the frequency extension modules from Virginia Diodes, Inc. (VDI), that can generate and detect microwaves in the frequency range from 75 to 330\,GHz. A superconducting magnet system from Oxford Instruments is used to produce uniform magnetic field up to 16\,T on the sample in a continuous field-sweep mode. The sample was mounted on a probehead equipped with oversized waveguides, which support broadband microwave propagation in transmission mode. $^4$He VTI inside the magnet system enabled precise control of temperatures between 3 and 300\,K.

Details of the primary analysis of the raw spectra are given in Appendix~\ref{section:analysis_spectra}.

\section{Structural and static magnetic properties}
\subsection{Powder x-ray diffraction}
\label{section:x-ray}
\begin{figure} 
    \includegraphics[width=\columnwidth]{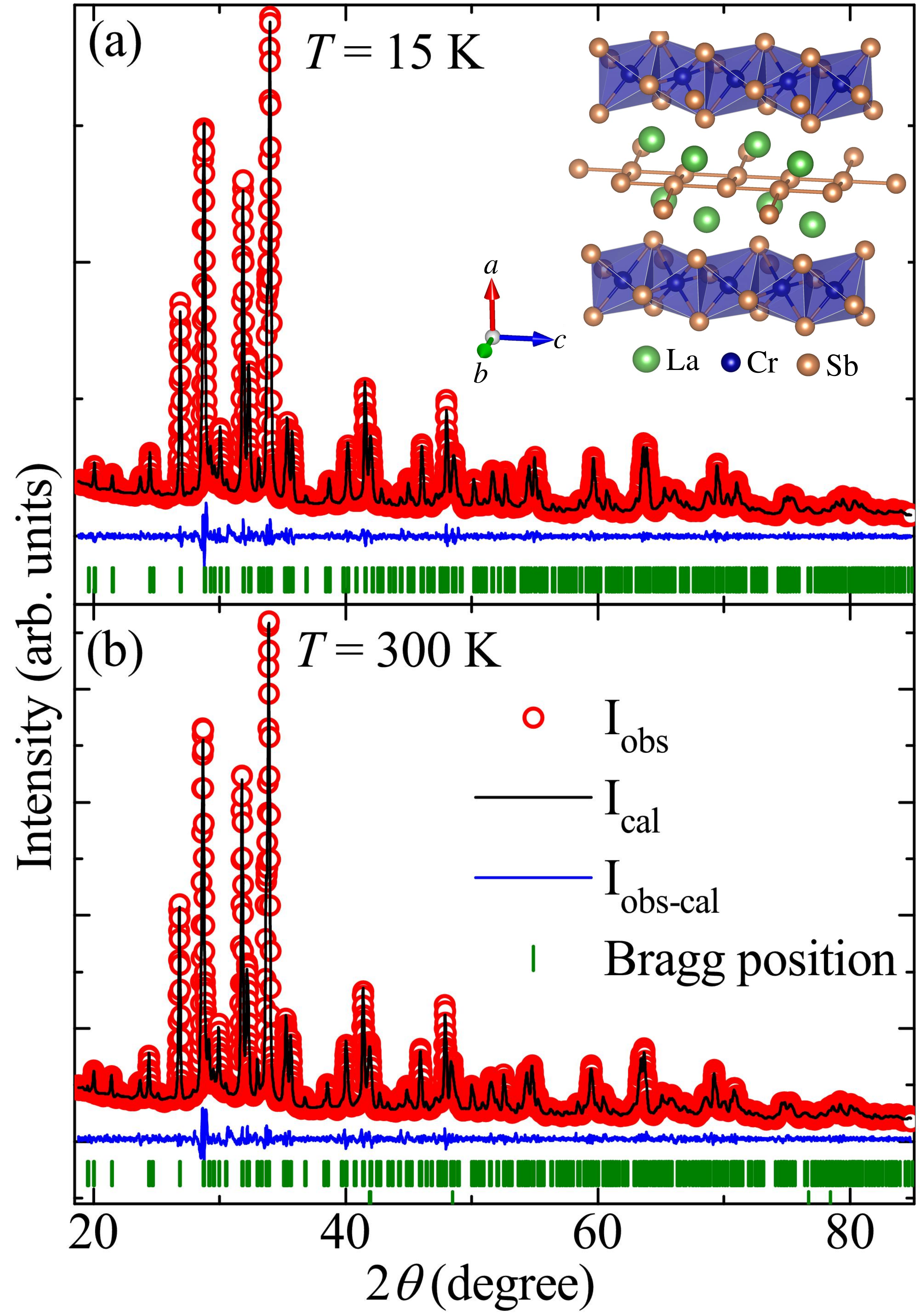}
    \caption{Powder XRD patterns (open red circles) at (a) $T = 15$~K and (b) $T = 300$~K. The solid black line indicates the Rietveld fit. The Bragg peak positions are shown by green vertical bars, and the bottom blue solid line represents the difference between the experimental and calculated intensities. Inset of (a) projects the crystal structure, highlighting the stacking sequence of the CrSb$_2$ layers ($bc$-plane) along the $a$-direction, which are well separated by La and Sb atoms.}
    \label{Fig1}
\end{figure}
\begin{figure}
	\includegraphics[width=\columnwidth]{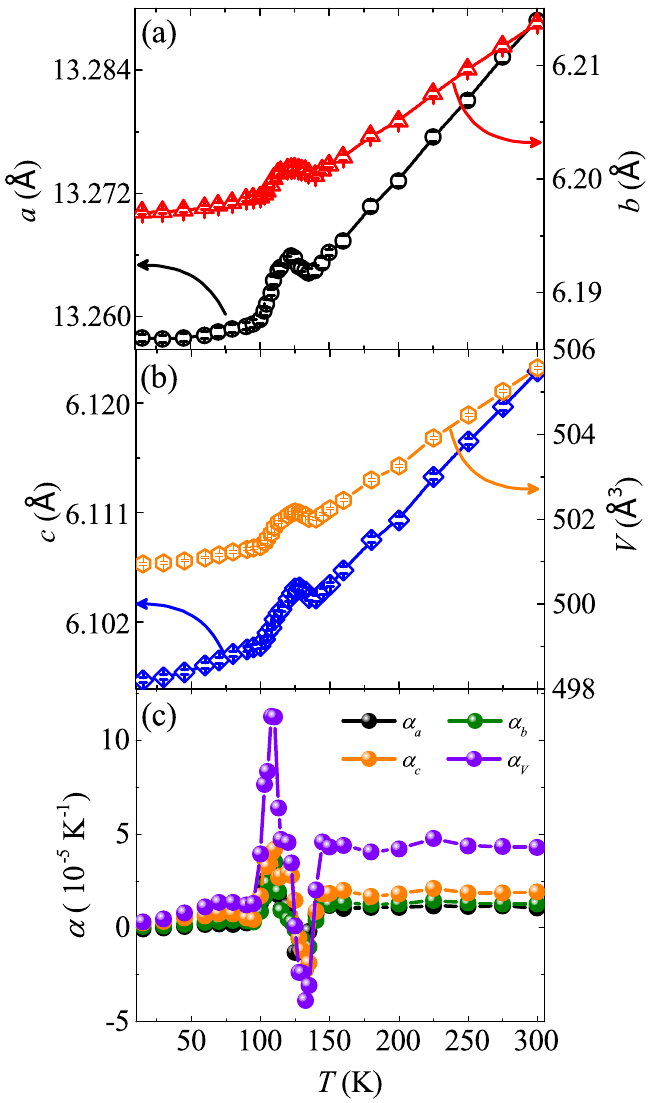}
	\caption{Temperature variation of lattice parameters (a) $a$ and $b$ and (b) $c$ and unit cell volume ($V_{\rm cell}$). (c) Thermal expansion coefficients ($\alpha$) as a function of temperature.} 
	\label{Fig2} 
\end{figure}
\begin{figure}
	\includegraphics[width=\columnwidth]{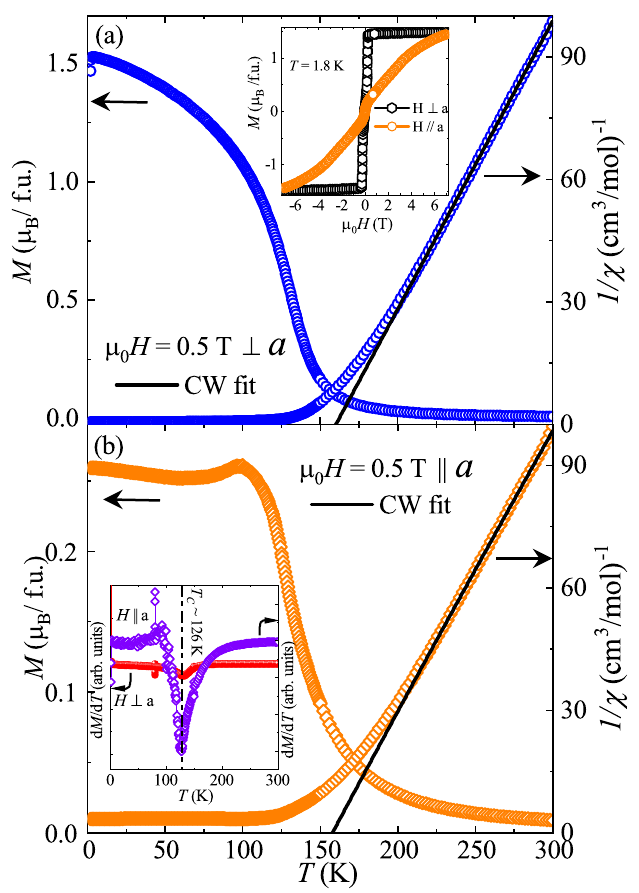}
	\caption{(a) Left $y$-axis: Temperature dependent dc magnetization $M$ measured in a magnetic field of $\mu_0H = 0.5$~T, applied perpendicular to $a$-axis. Right $y$-axis: $\chi^{-1}$ vs $T$ and the red solid line is the CW fit. Inset: Magnetic isotherm
		($M$ vs $H$) measured at $T = 1.8$~K along two diffident orientations. (b) Left $y$-axis: $M(T)$ measured at $\mu_0H=0.5$~T, parallel to the $a$-axis. Right $y$-axis: $\chi^{-1}$ vs $T$ along with the CW fit. Inset: $dM/dT$ vs $T$ for $H \perp a$ (left $y$-axis) and $H \parallel a$ (right $y$-axis), respectively.}
	\label{Fig3} 
\end{figure}
Figures~\ref{Fig1}(a) and (b) present the Rietveld refinement of the powder XRD patterns of LaCrSb$_3$ at $T = 15$~K and 300~K, respectively. Both the powder patterns are well indexed using the orthorhombic structure with space group $Pbcm$. The refined lattice parameters [$a = 13.288(3)~\text{\AA}$, $b = 6.213(8)~\text{\AA}$, $c = 6.122(6)~\text{\AA}$, and unit cell volume $V = 505.5(2)~\text{\AA}^3$] are in good agreement with the previous report~\cite{Ferguson191-198}. The temperature dependence of lattice parameters ($a$, $b$, $c$, and $V$) are depicted in Figs.~\ref{Fig2}(a-b). Upon cooling from 300~K, $a$, $b$, $c$, and $V$ decrease systematically, which is associated with the typical thermal contraction of the unit cell. As the temperature is lowered below $\sim 140$~K, all of them exhibit a slight increase, with a subtle anomaly at around $T_t \simeq 125$~K, where the magnetic transition sets in. This indicates a strong coupling between structural and magnetic degrees of freedom, recognized as magnetoelastic coupling. Similar results are reported in some other materials having a magneto-structural transition, e.g., in the mixed complex oxides (Ni,Co)TiO$_3$ and the van der Waals magnets (Fe,Ni)P$_2$S$_6$~\cite{Dey195122,Hoffmann014429,BesthaL020409}. Though no structural transition is detected for LaCrSb$_3$ down to 15~K, the observed anomaly at $T_t$ indicates either a change in symmetry or lattice distortion~\cite{Singh6981}.

To further investigate the connection between lattice expansion and magnetic properties, we evaluated the thermal expansion coefficients, defined as $\alpha_{\rm A} = \frac{1}{A}(\frac{dA}{dT})_P$, where $A$ stands for the lattice parameters ($a$, $b$, $c$, and $V$) and $P$ is the pressure~\cite{Singh033902}. The extracted thermal expansion coefficients are presented in Fig.~\ref{Fig2}(c) as a function of temperature. At the magnetic transition temperature $T_{\rm C} \simeq 126$\,K, all the thermal expansion coefficients exhibit a pronounced anomaly of a similar kind. This observation reveals that the magnetic transition is accompanied by an isotropic change in the unit cell. The minimum value of $\alpha_V$ is about $\sim -3.85 \times 10^{-5}$~K$^{-1}$ near $T_{\rm C}$, and it is comparable to the values reported for other materials having negative thermal expansion (NTE)~\cite{Song4690,Huang11469--11472,Takenaka131904}. 
Moreover, the sharp NTE peak at the magnetic transition temperature suggests high quality of the single crystals~\cite{Song4690}.

In the case of a second-order transition, the volumetric thermal expansion coefficient ($\alpha_V$) features a discontinuity at the transition, which is directly connected to the pressure dependence of the transition temperature $T_{\rm c}$ via the Ehrenfest relation,
\begin{equation}\label{Eq1}
    \frac{dT_{\rm c}}{dP} = \frac{\Delta\alpha_{\rm V} V_{\rm mol} T_{\rm C}}{\Delta C_{\rm p}},~\quad \text{near}~P=0.
\end{equation} 
Here, $\Delta \alpha_V$ is the volumetric thermal expansion coefficient, $V_{\rm mol}$ denotes the molar volume, and $\Delta C_{\rm p}$ represents the change in heat capacity at a constant pressure at $T_{\rm C}$.
The observed negative $\Delta\alpha_{\rm V}$ at $T_{\rm C}$ suggests, according to Eq.~(\ref{Eq1}), a negative $\frac{dT_{\rm c}}{dP}$, indicating that the transition temperature decreases with increasing external pressure, as observed in few other compounds~\cite{Singh033902,Eich096118}.

\begin{figure*}
	\centering
	\includegraphics[width=\linewidth]{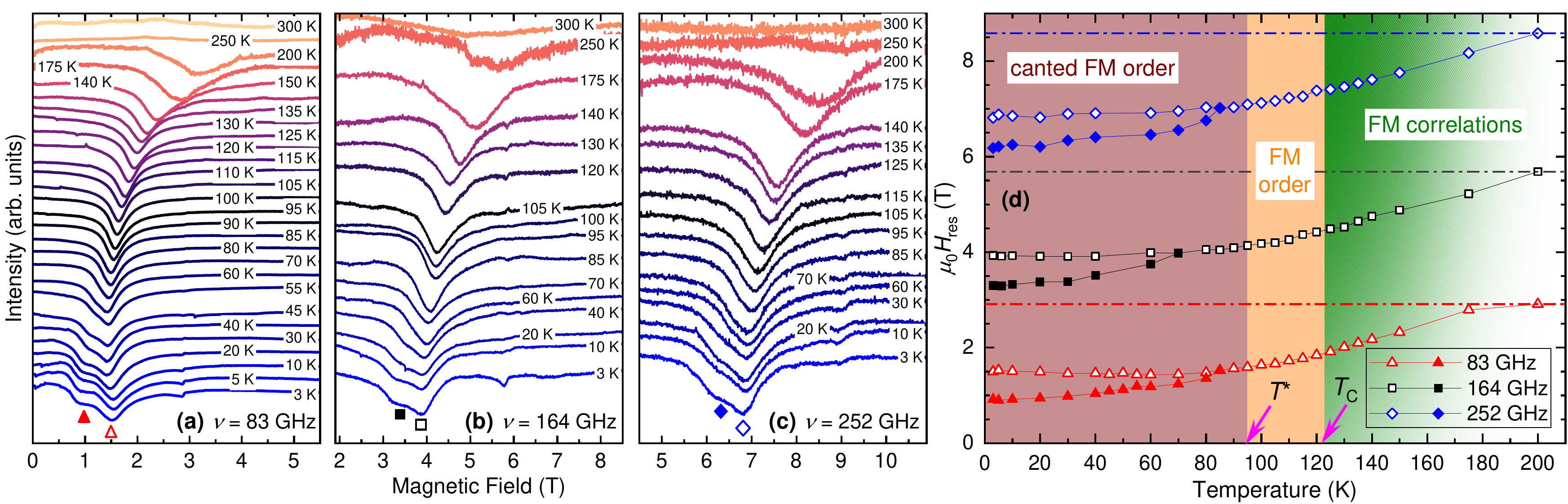}
	\caption{Temperature evolution of the HF-ESR powder spectrum of \LCS at a fixed microwave frequency of (a) 83\,GHz, (b) 164\,GHz, and (c) 252\,GHz. The spectral shape was corrected to obtain the pure absorption component of the detected signal as explained in Appendix~\ref{section:analysis_spectra}. The spectra are also normalized and shifted vertically for clarity. Symbols (same as in panel (d)) mark the position of the peak and of the shoulder of the spectra. (d) Temperature dependence of the resonance position $H_\mathrm{res}$ of the peak (open symbols) and the shoulder (closed symbols) of the powder spectrum at 83, 164 and 252\,GHz. Lines connecting data points are guides for the eye. The horizontal dashed line of the corresponding color denotes the expected paramagnetic position of the HF-ESR signal  at the respective frequencies according to Eq.~(\ref{eq:PM}). Color shaded regions correspond to reported magnetic phases in \LCS. The ordering temperature $T_{\rm C}$ and the temperature of the spin-reorientation transition $T^\ast$ are indicated by arrows. (see the text for details) }
	\label{fig:TDep}
\end{figure*}

\subsection{Magnetization}
\label{section:magnetization}

Temperature variation of magnetization $M(T)$ measured in an applied magnetic field of $\mu_0H = 0.5$~T perpendicular and parallel to the $a$-axis is presented in Fig.~\ref{Fig3}(a) and (b), respectively. In the high temperature regime, paramagnetic behavior is observed down to $T \sim 200$\,K. Below this temperature, $M(T)$ shows a large increase, suggesting the onset of a ferromagnetic (FM) transition. From the derivative $dM/dT$ shown in the inset of Fig.~\ref{Fig3}(b), the transition temperature is found to be $T_{\rm C} \simeq 126$~K\, which corresponds to the temperature of the anomaly $T_t$ seen in the temperature dependence of the lattice parameters. Moreover, the absolute value and shape of the $M(T)$ curves below $T_{\rm C}$ are distinctly different along different crystallographic orientations, which is an indication of strong magnetic anisotropy. The FM transition temperature is slightly different in other reports which could be due to different sample synthesis conditions~\cite {Jackson014421,Raju3630,Leonard4759}.

In the high temperature regime ($T > 235$~K), the inverse magnetic susceptibility ($\chi^{-1}$) measured at $\mu_0H = 0.5$~T was fitted by the Curie-Weiss (CW) law $\chi (T)=\chi_0 + C/(T-\theta_{\rm CW})$ (right $y$-axes in Fig.~\ref{Fig3}). Here, $\chi_0$ represents the $T$-independent susceptibility, $C$ is the Curie constant, and $\theta_{\rm CW}$ is the CW temperature that represents the average interaction among the magnetic entities. The fit returns, $\chi_0 \simeq 2.81(3)\times 10^{-4}$~cm$^3$/mol, $C \simeq 1.41(1) $~cm$^3$K/mol, and $\theta_{\rm CW} \simeq 160.2(5)$~K and $\chi_0 \simeq 2.64(2) \times 10^{-4}$~cm$^3$/mol, $C \simeq 1.40(2)$~cm$^3$K/mol, and $\theta_{\rm CW} \simeq 157.8(4)$~K for $H \perp a$ and $H \parallel a$, respectively. From the values of $C$, the effective magnetic moment is calculated to be $\mu_{\rm eff} \simeq 3.35$~$\mu_{\rm B}$ and $\sim 3.34$~$\mu_{\rm B}$, respectively. These values of $\mu_{\rm eff}$ are slightly smaller as compared to the expected theoretical value 3.87~$\mu_{\rm B}$ for free Cr$^{3+}$ ions, suggesting the itinerant character of the compound. The positive value of $\theta_{\rm CW}$ indicates the presence of predominant FM interactions among the Cr spins.

The magnetic isotherm ($M$ vs $H$) measured at $T = 1.8$~K in two different crystallographic directions is displayed in the inset of Fig.~\ref{Fig3}(a). Magnetization for $H \perp a$ direction saturates to a value of $\sim 1.5$~$\mu_{\rm B}$ in a very small applied field, whereas, for the $H \parallel a$ direction, saturation is not reached even at 7~T. This indicates strong magnetic anisotropy with the $bc$-plane being the easy plane. Moreover, a step-like feature is observed in a small field of $\sim 0.2$~T for both directions, suggesting a field-induced meta-magnetic transition~\cite{Jackson014421}. All these observations as well as additional $M(H)$ data shown in Appendix~\ref{section:additional_data} are consistent with the previous reports~\cite{Jackson014421,Yang024419}.

\section{ESR/FMR spectroscopy}
Though technically the same equipment is used for measuring ESR and FMR, one should note the distinction between these two phenomena. The former is the resonance excitation of paramagnetic, possibly exchange coupled spins, whereas the latter is the collective resonance excitation of the ordered spin lattice -- spin waves or magnons. Since the wavelength of the applied microwave radiation is much larger than the interatomic distances, only the uniform wavevector $\mathbf{q} \rightarrow 0 $ mode can be probed in this experiment. In the simplest case of a one sublattice ferromagnet it corresponds to the resonance oscillation of the total magnetization vector of the sample. Three spectrometers used in this work, HF-ESR (75 -- 330\,GHz), frequency-swept ESR (1 -- 67\,GHz), and X-band ESR (9.56\,GHz), cover a broad range of frequencies and corresponding magnetic fields providing a comprehensive picture of FMR excitations in \LCS. Due to the limited sensitivity of the HF-ESR setup additionally complicated by the low penetration depth of the high-frequency microwaves in metallic \LCS, only measurements on the powder sample were possible, whereas the other two more sensitive setups operational at lower frequencies enable measurements on single crystals.

\subsection{Frequency range 75--330\,GHz}

HF-ESR spectra were measured on the powder sample of \LCS  in the temperature range of 3 -- 300\,K at three different microwave frequencies as shown in Fig~\ref{fig:TDep}(a)--(c). The signal emerges first as a very broad absorption line at $T \lesssim 250$\,K, and then narrows and shifts to lower fields with decreasing the temperature. Remarkably, at $T < 80$\,K, a shoulder develops on the left side  of the main peak, which becomes prominent at lower temperatures. The temperature dependence of the resonance field of the main peak and the shoulder at three studied excitation frequencies is plotted in Fig.~\ref{fig:TDep}(d). The horizontal lines in this plot denote the  expected paramagnetic resonance field $H^{\rm par}_{\rm res}$ at a given frequency which  is calculated using the conventional ESR paramagnetic relation \cite{Abragam2012}:
\begin{equation}
	h\nu = g\mu_{\rm B}\mu_{0}H^{\rm par}_{\rm res}(200\,{\rm K}),
	\label{eq:PM}
\end{equation}
where $h\nu$ represents the microwave energy provided to the sample, $\mu_{\rm B}$ is the Bohr magneton, $\mu_0$ is the permeability of the free space and $g$ is the $g$-factor of resonating spins. The $g$-factor is obtained from the frequency dependence of the resonance field of the signal at $T = 200$\,K, as explained below. The shift of the powder spectrum from $H^{\rm par}_{\rm res}$  is clearly visible below $T$ $\sim$ 200\,K, i.e., well above $T_{\rm C}$, suggesting a gradual development of the static on the ESR timescale short range correlations in \LCS. 
 
\begin{figure}
	\centering
	\includegraphics[width=\linewidth]{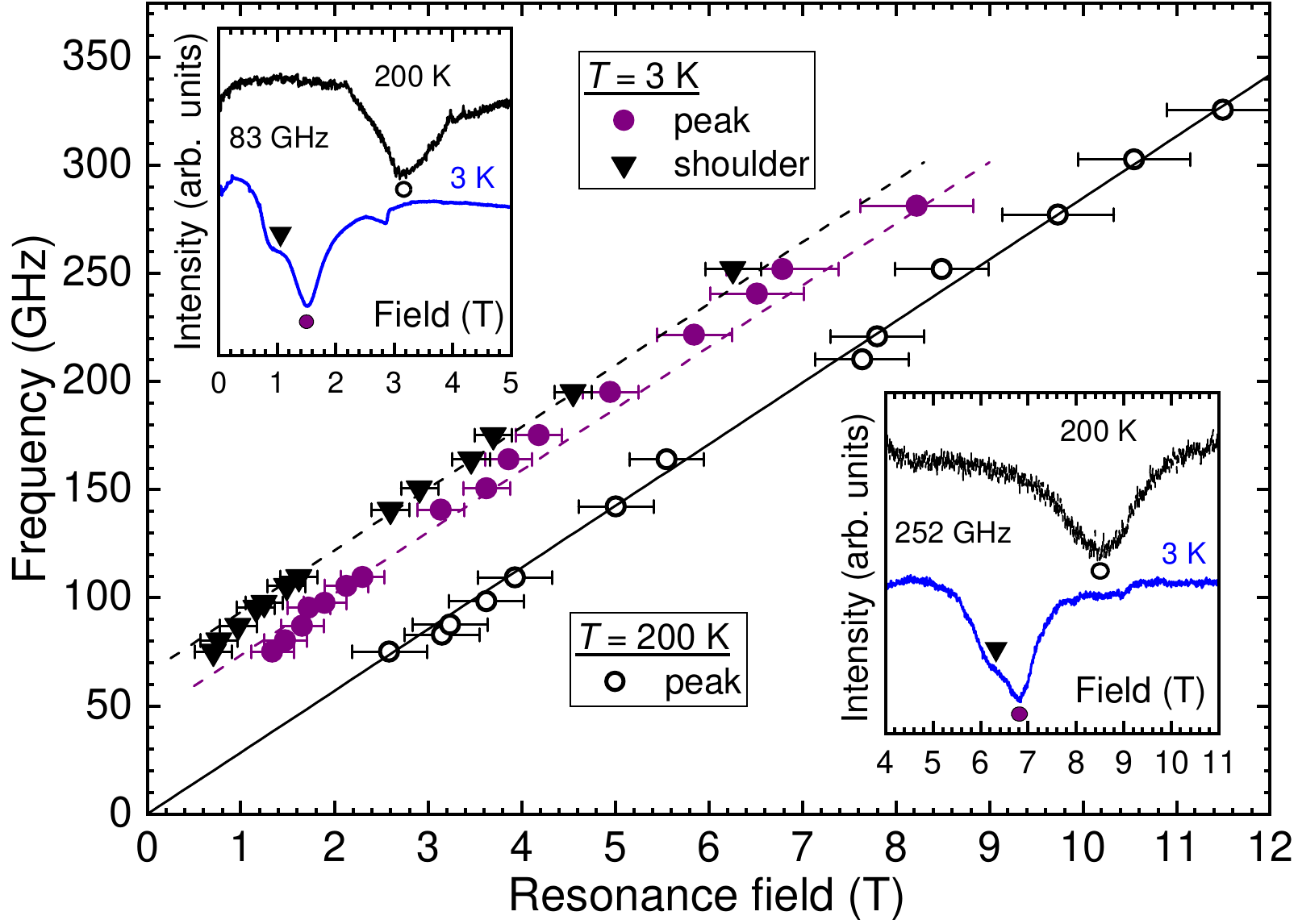}
	\caption{Main panel: Frequency-field relation $\nu(H_{\rm res})$ of the resonance signals  of the powder sample of \LCS at 200\,K (open circles) and at 3\,K (closed circles - main peak, closed triangles - shoulder). Solid straight line is the fit to the paramagnetic resonance condition according to Eq.~(\ref{eq:PM}). Dashed lines are guides for the eye. Insets: Representative ESR  spectra at two selected frequencies of 83 and 252\,GHz recorded at 3 and 200\,K. Symbols (same as in the main panel) indicate the position of the main peak and the shoulder of the powder spectrum of \LCS.}
	\label{fig:FDep_200K_3K}
\end{figure}

The frequency-resonance field dependence $\nu(H_{\rm res})$ of the HF-ESR signal of the powder sample of \LCS measured in the paramagnetic state at $T=200$\,K in the frequency range 75--330\,GHz is plotted in Fig.~\ref{fig:FDep_200K_3K} by open circles. A linear dependence is observed, which can be fitted with the conventional paramagnetic resonance condition for a gapless excitation, as given by Eq.~(\ref{eq:PM}). The fit yields the powder averaged $g$-factor, $g=2.04\pm0.16$, close to the spin-only value of 2. This value, within the error bars, is consistent with the theoretical predictions for a Cr$^{3+}$ ion in the high-spin state~\cite{Abragam2012}. 

Below $T_{\rm C}$, the measured signal corresponds to the FMR response of the ordered spin lattice of \LCS.  The $\nu(H_{\rm res})$ dependence of the main peak and the shoulder of the low-temperature powder FMR spectrum measured at $T = 3$\,K in the same frequency range is shown in Fig.~\ref{fig:FDep_200K_3K} by closed circles and triangles, respectively. It features a significant up-frequency offset from the paramagnetic resonance branch.

\subsection{Frequency range 1--67\,GHz}
\begin{figure*}
	\includegraphics[width=\linewidth]{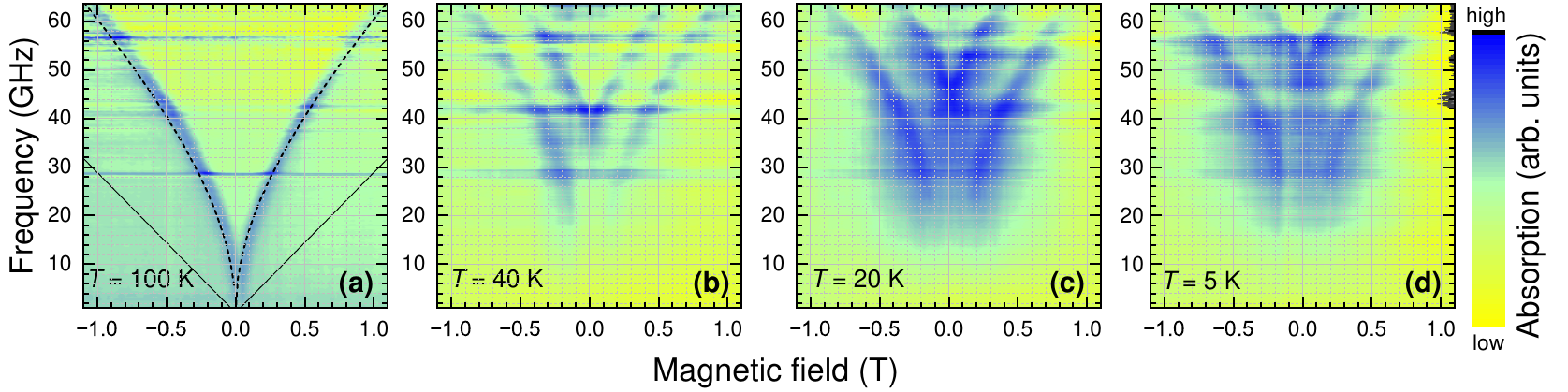}
	\caption{Two-dimensional color-coded view of the frequency swept FMR spectra of a single crystal of \LCS for $\mathbf{H}\parallel \mathbf{c}$ field geometry at (a) 100\,K, (b) 40\,K, (c) 20\,K, and (d) 5\,K. Dashed curve in (a) is the fit of the data to Eq.~(\ref{eq:LWST1}), and the thin solid line represents the paramagnetic branch according to Eq.~(\ref{eq:PM}). Horizontal stripes are artefacts caused by parasitic resonances in the coplanar waveguide. }
	\label{fig:coplanarCaxis}
\end{figure*}

To investigate the low-frequency part of the $\nu(H_{\rm res})$ diagram, frequency-swept spectra in the range 1 -- 67\,GHz were recorded with the coplanar-waveguide ESR setup. Due to its higher sensitivity as compared to the HF-ESR spectrometer, measurements on a single crystalline sample of \LCS were possible. Representative FMR spectra at selected temperatures of 5, 20, 40, and 100\,K for the magnetic field applied along the $c$-axis of the crystal are presented in Fig.~\ref{fig:coplanarCaxis} and those for $\mathbf{H}\parallel \mathbf{b}$ are shown in Appendix~\ref{section:additional_data}, Fig.~\ref{fig:coplanarBaxis}.    

The FMR response at 100\,K features a single non-linear gapless resonance branch typical for a one sublattice uniaxial  easy-plane ferromagnet with the magnetic field vector lying in the easy plane. It significantly deviates from the linear paramagnetic branch calculated according to Eq.~(\ref{eq:PM}), as shown in Fig.~\ref{fig:coplanarCaxis}(a) by the thin solid line (see also Fig.~\ref{fig:FDep_200K_3K}). 
Indeed, the observed non-linear branch  can be very well fitted with the analytical solution in the frame of LSWT for the easy plane field geometry \cite{Turov}: 
\begin{equation}
	h\nu = g\mu_{\rm B}\mu_{0} \sqrt{H(H+ \arrowvert H_{\rm a}\arrowvert)}\, .
\label{eq:LWST1}
\end{equation}
Here,  $H_{\rm a}$ is the effective anisotropy field  amounting to 3.65\,T that forces the spins to lie in the $bc$-plane (dashed curve in Fig.~\ref{fig:coplanarCaxis}(a) and Fig.~\ref{fig:coplanarBaxis}(a)). In contrast, the excitation spectrum at lower temperatures becomes more complex. The absorption intensity below 20\,GHz strongly diminishes and the second branch up-shifted from the first one  becomes clearly visible in the spectrum (Fig.~\ref{fig:coplanarCaxis}(b)-(d)). The lower and the higher frequency branches are in apparent correspondence with the peak and the shoulder of the low-temperature FMR powder spectrum (cf.~Fig.~\ref{fig:FDep_200K_3K}). Correspondingly, the magnetization measured as a function of field for $\mathbf{H}\perp \mathbf{a}$ at $T = 110$\,K does not exhibit any signatures of metamagnetic transition, as expected for an easy-plane ferromagnet, whereas at low temperatures it demonstrates  a more complex behavior indicating the presence of an additional easy axis in the $bc$-plane (Appendix~\ref{section:additional_data}, Fig.~\ref{fig:MH}). 

Similar evolution of the FMR response is observed also for the $\mathbf{H}\parallel \mathbf{b}$ field geometry (Fig.~\ref{fig:coplanarBaxis}(b)-(d) in Appendix~\ref{section:additional_data}). The possible origin and modeling of the frequency-field dependence of these two branches will be discussed in detail in Sect.~\ref{section:model}.

\begin{figure*}
	\centering
	\includegraphics[width=\linewidth]{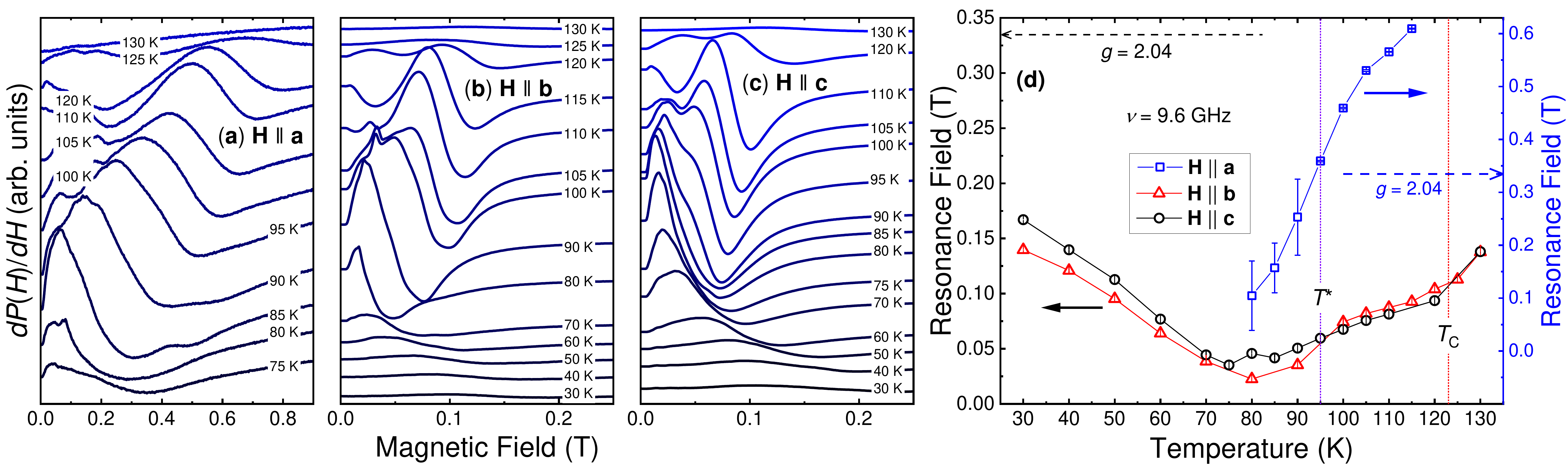}
	\caption{Temperature dependence of the FMR spectra (field derivatives of the microwave absorption $dP(H)/dH$) measured at the X-band frequency of 9.56\,GHz on a single crystal of \LCS with the applied field along (a) \HIIa, (b)\HIIb, and (c) \HIIc directions. (d) Temperature dependence of the resonance field of the signals $H_{\rm res}(T)$ extracted from the spectra shown in (a)--(c). Left vertical axis: \HIIb and \HIIc field geometries. Right vertical axis: \HIIa orientation. Horizontal dashed arrows denote the expected paramagnetic position of the signal at this frequency according to Eq.~(\ref{eq:PM}). The vertical dotted lines indicate the FM ordering temperature $T_{\rm C}$ and the spin-reorientation transition temperature $T^\ast$.}
	\label{fig:XBand}
\end{figure*}

\subsection{X-band ESR at 9.56\,GHz}\label{sec:results_XBand}
Finally, the temperature-dependent ESR spectra of the single-crystalline sample of \LCS were recorded at a fixed X-band frequency of 9.56\,GHz, with the external magnetic field applied along the \HIIa, \HIIb, and \HIIc crystallographic directions, as shown in Fig.~\ref{fig:XBand}(a)-(c). In all configurations, the resonance lines emerge upon cooling below 130\,K, i.e., close to the FM ordering temperature $T_{\rm C}$, and thus can be considered as FMR signals. For \HIIb and \HIIc, the spectrum consists of the main resonance line and a smaller signal on its low-field side. Both gradually shift towards lower magnetic fields with decreasing temperature and eventually merge into one broad resonance line. This small signal is probably not related to FMR but arises from breaking of domain walls during the field sweep causing a non-resonant microwave absorption. Below 80\,K, the FMR signal shifts toward higher fields, progressively broadening and eventually flattening around 30\,K. For \HIIa, the spectrum consists of a single broad line which rapidly shifts toward low fields and eventually disappears around 80\,K.

The corresponding temperature dependence of the resonance field $H_{\rm res}(T)$ for all three measured field directions is shown in Fig.~\ref{fig:XBand}(d). The signals are strongly shifted from the expected paramagnetic position denoted in this plot by horizontal dashed arrows. This is similar to the line shifts observed in the FM ordered state at higher frequencies (Fig.~\ref{fig:TDep} and Fig.~\ref{fig:coplanarCaxis}). Interestingly, below the spin reorientation transition $T^\ast$ the signal intensity rapidly wipes out as it is also the case for the frequency swept FMR spectra at this low frequency (Fig.~\ref{fig:coplanarCaxis}(b)-(d)).  

\section{Discussion}

\subsection{Short-range ordered state state above $\boldsymbol{T_{\rm C}}$}

It is remarkable that the ESR signal from \LCS is not visible at room temperature, i.e., in the truly paramagnetic state of this compound, but emerges by approaching the FM ordering temperature $T_{\rm C} \backsimeq 126$\,K from above. ESR in the paramagnetic state implies the coherent precession of individual magnetic moments which in metallic systems can be rapidly destroyed by the fast relaxation processes within the bath of conduction electrons \cite{Barnes1981}. In contrast, FMR is the collective resonance of the entire ordered spin lattice which is much less affected by destructive relaxation effects. Therefore, the appearance of the signal by approaching $T_{\rm C}$ suggests the development of the  spin-spin correlations in the system that are quasi-static on the fast nanosecond ESR timescale. They give rise to the quasi-static internal field which shifts the resonance line from its paramagnetic position. Such shifts above ordering temperature are commonly observed in layered materials with weak interlayer exchange, where reduced dimensionality enhances two-dimensional (2D) spin correlations preceding the 3D order~\cite{Kataev2024, Abraham2023, Magar2022, Zeisner2020, Abraham2025}.     
The magnitude of the resonance shift decreases at lower frequencies (cf. Fig.~\ref{fig:TDep}(d) and Fig.~\ref{fig:XBand}(d)), indicating a field-induced enhancement of short-range correlations, as expected for the FM type of interactions~\cite{Zeisner2019, Abraham2025}. These findings are consistent with the positive Curie--Weiss temperature ($\theta_{\rm CW}$ $>$ $T_\mathrm{C}$) obtained from the temperature-dependent magnetization. The presence of such pronounced short-range correlations in \LCS is also corroborated by heat-capacity measurements~\cite{Granado107204, Inamdar2009} and recent analysis of the static magnetization \cite{Mao2026}. The absence of a clear $\lambda$-type anomaly at the FM transition, despite the large ordered moments, points to a gradual release of magnetic entropy over a broad range of temperatures due to the residual spin--spin correlations that persist above $T_\mathrm{C}$.

\subsection{Long-range ordered state at  $\boldsymbol{T < T_{\rm C}}$}

As the material is cooled down below $T_\mathrm{C}\simeq 126$\,K, the long-range  FM order sets in, with the hard magnetic axis along the crystallographic $a$-axis and the easy-plane ($bc$-plane) normal to it. 
According to the x-ray diffraction data, all three thermal expansion coefficients $\alpha_{\rm a}$, $\alpha_{\rm b}$, and $\alpha_{\rm c}$, exhibit a pronounced anomaly at $T_{\rm C}$, indicating a rather isotropic change of the unit cell volume (see Sect.~\ref{section:x-ray}). However, it appears that this change has no visible effect on the properties of FMR. In particular, no anomaly can be discerned within the experimental error bars  in the temperature dependence of the resonance field at $T_{\rm C}$ (Fig.~\ref{fig:TDep}(d)), suggesting that an isotropic change of the lattice parameters does not significantly affect magnetic anisotropy which defines the position of the FMR signal.    

Below $T_{\rm C}$, there is a small anisotropy in the easy plane aligning the Cr spins toward the $c$-axis with an alternating canting of the spins away from this axis by about $18^\circ$ \cite{Granado107204}. However, this anisotropy is such weak that the application of a small field of 40\,mT along the $b$-axis at $T = 100$\,K turns the spins in this direction \cite{Granado107204}. This value gives the strength of the additional internal in-plane anisotropy field $H_{\rm a}^{\rm c}$ which keeps the spins in the $c$-axis direction in the absence of external field. This field should open a gap in the FMR excitation spectrum at $H = 0$ amounting approximately to $H_{\rm a}^{\rm c}(g\mu_{\rm B}\mu_{0}/h) \sim 1.2$\,GHz, only \cite{Turov}. This is consistent with the observation of the practically gapless FMR branch for \HIIc (Fig.~\ref{fig:coplanarCaxis}(a) and Fig.~\ref{fig:coplanarBaxis}(a)) and the occurrence of the FMR signals in the low-field region at $\nu = 9.56$\,GHz (Fig.~\ref{fig:XBand}). Note, that $H_{\rm a}^{\rm c}$ is much smaller than the anisotropy field $H_{\rm a} = 3.65$\,T responsible for the in-plane alignment of the ordered spins that was estimated from the fit of the FMR branch at $T = 100$\,K to Eq.~(\ref{eq:LWST1}) (Fig.~\ref{fig:coplanarCaxis}(a) and Fig.~\ref{fig:coplanarBaxis}(a)).  

On further cooling down below the spin-reorientation transition $T^\ast \sim 95$\,K, the spins turn toward the $b$-axis direction \cite{Jackson014421,Granado107204,Yang024419} while retaining the alternating canting of the spins by $18^\circ$ \cite{Granado107204}. According to the x-ray diffraction data, there is no lattice anomaly at $T^\ast$, suggesting the  electronic nature of the spin reorientation. This newly emergent $b$-axis anisotropy at $T < T^\ast$  is significantly stronger as compared to the $c$-axis anisotropy that was present at $T > T^\ast$. This is evident from the depletion of the FMR absorption intensity around the zero field (Fig.~\ref{fig:coplanarCaxis}(b)--(d)), and rapid shifts and disappearance of the FMR signals at $\nu = 9.56$\,GHz (Fig.~\ref{fig:XBand}), suggesting  an opening of an energy gap for the FMR excitations at $H =0$. Note, that FMR excitations of an easy-plane ferromagnet without an additional in-plane anisotropy would remain gapless \cite{Turov}. 

\begin{figure}
	\centering
	\includegraphics[width=\linewidth]{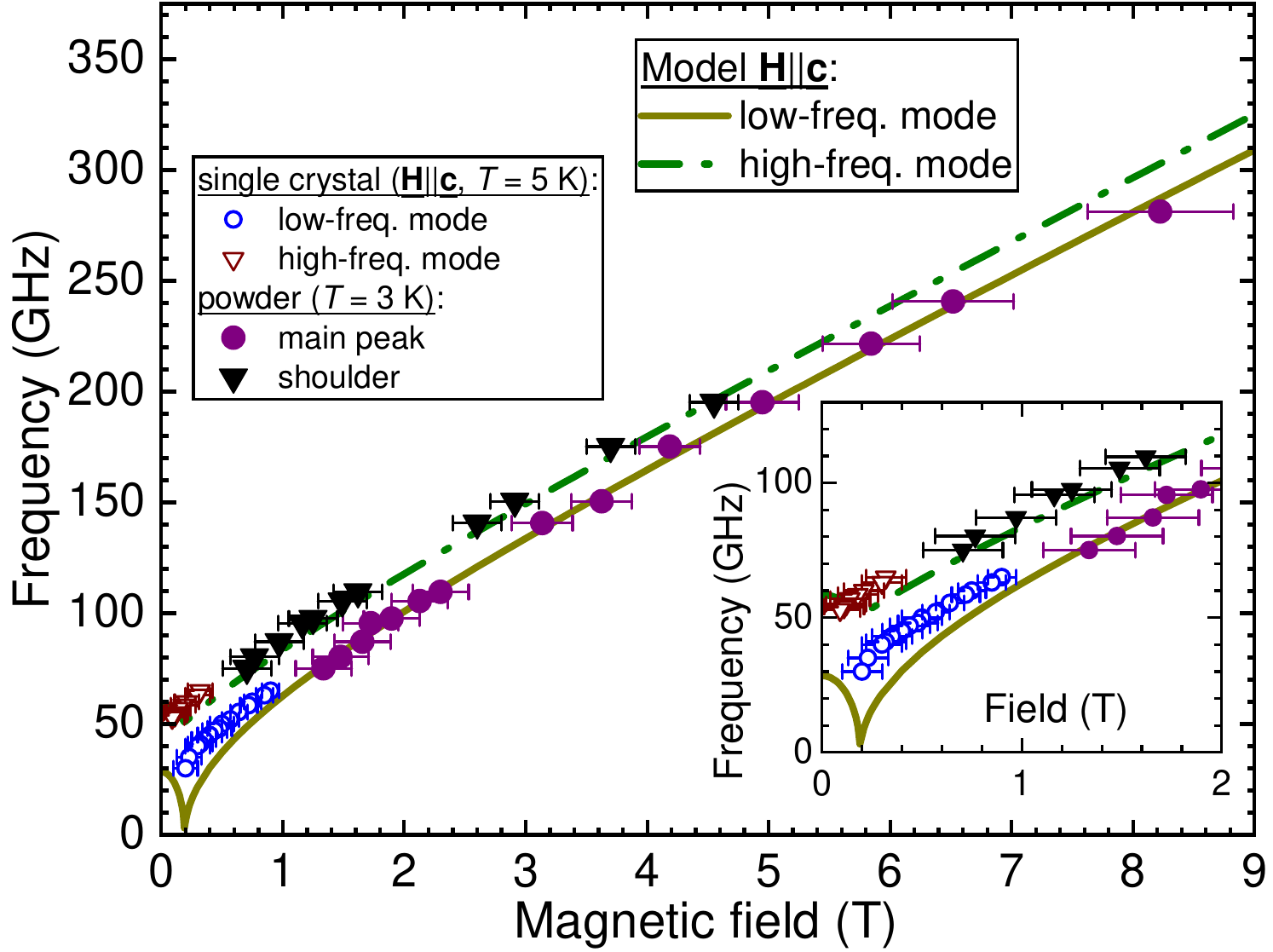}
	\caption{Summary $\nu(H_{\rm res})$ diagram of the experimentally determined FMR branches. Solid circles and triangles are the resonance fields of the peak and the shoulder of the FMR powder spectrum taken from Fig.~\ref{fig:FDep_200K_3K}. Open circles and triangles are the single crystalline data extracted from the frequency swept spectra in Fig.~\ref{fig:coplanarCaxis}.
	Solid and dash-dot lines are the results of the modeling for $\mathbf{H}\parallel \mathbf{c}$ discussed in Sect.~\ref{section:model} using parameters listed in Table~\ref{tab:model_param}.  Inset: enlarged low-field part of the frequency-field diagram.}
	\label{fig:model_branches}
\end{figure}

The central result of the FMR experiments is the observation of the splitting of the FMR branches at $T \ll T^\ast$ (Fig.~\ref{fig:coplanarCaxis} and Fig.~\ref{fig:coplanarBaxis}). Their $\nu(H_{\rm res})$ dependence extracted from the two-dimensional frequency swept FMR spectra at the lowest temperature (Fig.~\ref{fig:coplanarCaxis}(d)) is plotted in Fig.~\ref{fig:model_branches} together with the polycrystalline FMR data. These two data sets match remarkably well suggesting that the two branches clearly resolved in the single crystalline FMR spectra average out into a peak and a shoulder of the corresponding powder spectra. Their possible origin and relation to the unusual canted FM magnetic order in \LCS will be discussed in the next Section.   

\subsection{Phenomenological model of FMR excitations}
\label{section:model}

\begin{figure*}
	\centering
	\includegraphics[width=\linewidth]{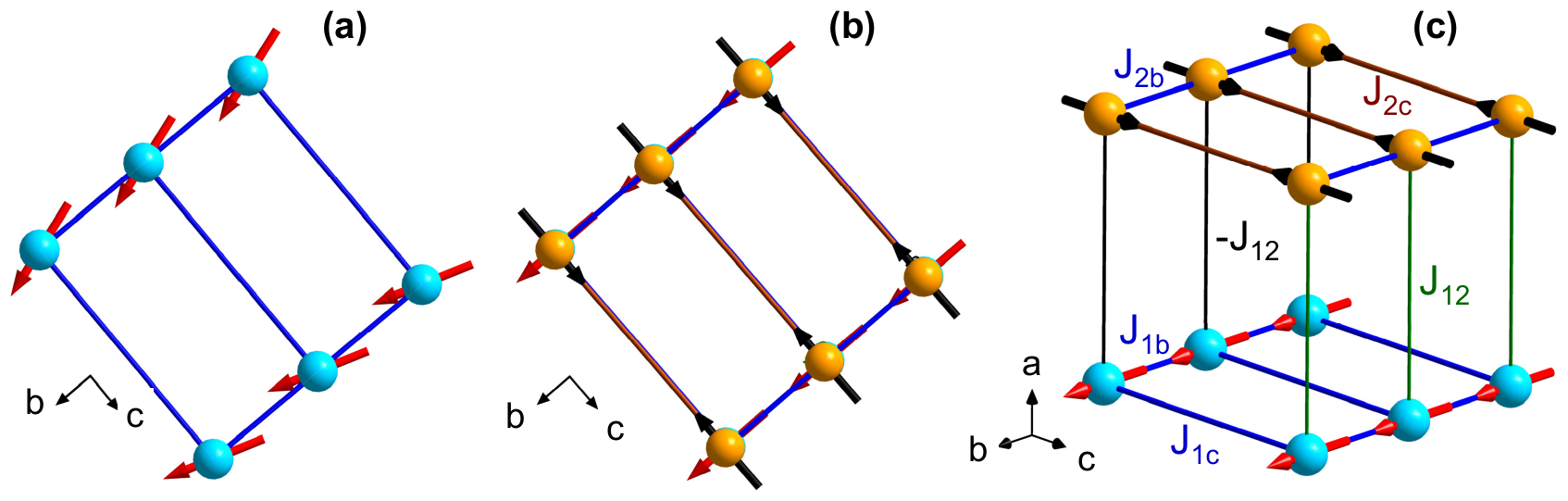}
	\caption{(a) In-plane view of the canted FM spin structure of \LCS observed by neutron diffraction \cite{Granado107204}. The spins are tilted off the easy-axis ($b$-axis) by $\pm 18\degree$. The tilting alternates along the $c$-axis. (b) Schematic of the FM and orthogonal AFM sublattices coexisting at the same crystallographic Cr sites in \LCS as proposed in Ref.~\cite{Granado107204}. (c) Intra- and inter-sublattice exchange paths with coupling matrices $J_{1b}, J_{1c}, J_{2b}, J_{2c}$ and $J_{12}$ considered for modeling of FMR modes [see Eqs.~(\ref{eq:J1matrix}), (\ref{eq:J2matrix}), (\ref{eq:J12matrix})]. For clarity, the sublattices are artificially  separated along the $a$-axis (for details see Sect.~\ref{section:model}).  }
	\label{fig:spin_structure}
\end{figure*}

The unit cell of \LCS contains 4 equivalent magnetic  Cr ions (Fig.~\ref{figure:structure}). Their site positions are related by the symmetry operations of the respective space group and they are expected to build up  a unique FM ordered lattice at $T < T_{\rm C}$. In this case, within the LSWT formalism, there should be four spin wave modes, three higher-in-energy "optical" branches, and one lowest-in-energy "acoustic" branch \cite{Prosnikov2022}. The latter one is often termed as transversal phase mode or pseudo--Goldstone mode. The acoustic mode is a collective in-phase uniform precession of the FM ordered spins, which in the limit of the wave-vector $\mathbf{q} = 0$ corresponds to the uniform precession of the total magnetization. The microwaves applied to the system couple to the transverse component of the magnetization and give rise to an FMR signal.
In contrast, the optical modes are the result of the anti-phase precession of the spins. Their transverse components may cancel out each other and produce no net transverse magnetization to which the microwaves can couple, rendering optical modes FMR-silent. Indeed, \LCS features such an acoustic FMR branch at temperatures slightly below $T_{\rm C}$ (Fig.~\ref{fig:coplanarCaxis}(a)). However, its splitting into two branches at lower temperatures    (Fig.~\ref{fig:coplanarCaxis}(b)-(d)) is not expected within the above framework.  

Observation of the two FMR branches in a single crystal of \LCS as well as the occurrence of a shoulder of the FMR signal from a powder sample, give strong indications that the magnetic lattice in \LCS consists of at least two distinct sub-units. Indeed, coexistence of FM and AFM sublattices  was proposed to explain the spin structure in the ordered state of \LCS inferred from neutron diffraction experiments \cite{Granado107204}. To rationalize the observed significant 18$\degree$ alternating canting of the FM ordered spins from the in-plane easy axis (Fig.~\ref{fig:spin_structure}(a)), it was conjectured that two species of magnetic electrons at the same Cr site in \LCS, one localized and the other one itinerant, with distinct exchange interactions and anisotropies build up an FM sublattice and an orthogonal to it AFM sublattice (Fig.~\ref{fig:spin_structure}(b)). Superposition of these sublattices  yields the observed canted FM spin structure. 
The FMR branches for the case of orthogonal and interacting FM and AFM sublattices  can be calculated with the SpinW software used for the analysis of the inelastic neutron scattering experiments \cite{Toth_2015,SpinW}. For that we restrict ourselves to a minimal phenomenological model which does not consider the specific microscopic origin of the two sublattices, since, from a theoretical perspective, the issue of duality of Cr $d$-electrons  is not settled so far. 
E.g., first-principle electronic structure calculations in Ref.~\cite{Richter2004} did not give a clue for the localized Cr electrons whereas {\it ab initio} calculation approach used in Ref.~\cite{Shim2004} favored coexistence of local and itinerant Cr $t_{\rm 2g}$ states in \LCS. 

\out{This} Our phenomenological model can be numerically solved with SpinW using LSWT. In SpinW, important input parameters are the exchange interaction tensors. The following set of exchange matrices was introduced for the calculations. In the FM sublattice the spins are coupled along the $b$- and $c$-axes with anisotropic FM tensors $J_{1b}$ and $J_{1c}$, respectively, (Fig.~\ref{fig:spin_structure}(c)) in the form:
\begin{equation}
	J_{1b} =
\begin{pmatrix}
	J^{xx}_{1b} & 0 & 0 \\
	 0 & J^{yy}_{1b} & 0\\
	 0 & 0 & J^{zz}_{1b} 
\end{pmatrix}
;
J_{1c} =
\begin{pmatrix}
	J^{xx}_{1c} & 0 & 0 \\
	0 & J^{yy}_{1c} & 0\\
	0 & 0 & J^{zz}_{1c} 
\end{pmatrix}
.
\label{eq:J1matrix}
\end{equation}
To ensure the easy-plane type of FM order with the magnetic easy $b$-axis  the following constraints were set: $J^{xx}_{1b} > J^{yy}_{1b} = J^{zz}_{1b}$, $J^{yy}_{1c} > J^{xx}_{1c} = J^{zz}_{1c}$ and $J^{xx}_{1b} > J^{yy}_{1c}$.

To achieve the alternating canting of the resulting spin structure along the $c$-axis (Fig.~\ref{fig:spin_structure}(a)), the AFM sublattice should be, in fact, of the stripe-type  order with FM coupling along the $b$-axis, $J_{2b} < 0$, and AFM coupling along the $c$-axis, $J_{2c} > 0$ (Fig.~\ref{fig:spin_structure}(c)). For simplicity, they are assumed to be isotropic and equal in magnitude: 
\begin{equation}
	J_{2b} = -J_{2c}\, . 
\label{eq:J2matrix}
\end{equation}

Finally, to keep the FM and AFM sublattices orthogonal the interlattice coupling tensor was set in the form:    
\begin{equation}
J_{12} =
\begin{pmatrix}
	0 & \pm J^{xy}_{12} & 0 \\
	\pm J^{yx}_{12} & 0 & 0\\
	0 & 0 & 0 
\end{pmatrix}
;
\ J^{xy}_{12} = J^{yx}_{12}\, .
\label{eq:J12matrix}
\end{equation}
Here, plus and minus signs of the off-diagonal components alternate from site to site along the $c$-axis (Fig.~\ref{fig:spin_structure}(c)). 

Further input parameters were the isotropic $g$-factor $g = 2$ obtained from the ESR measurements and the ordered moments of the FM and AFM sublattices 1.65 and 0.49$\mu_{\rm B}$/Cr, respectively, estimated in Ref.~\cite{Granado107204}.  

Unfortunately, not much is known so far about exchange interactions in \LCS. Merely, the FM Curie-Weiss constant $\Theta_{\rm CW} = 160$\,K obtained in the magnetization measurements (Sect.~\ref{section:magnetization}) suggests their average energy scale to be of the order of several meV. In this situation some ambiguity in setting the values of exchange parameters is unavoidable. However, it should be noted at this place, that the energies of the acoustic modes probed by ESR spectroscopy do not depend on the exchange constants but are determined  in zero magnetic field  by the geometrical mean of exchange $E_{\rm exch}$ and anisotropy $A$ energies $\sqrt{E_{\rm exch}A} $ for an antiferromagnet, and by $A$ alone for a ferromagnet \cite{Turov}. Therefore, the exchange tensors in the model can be initially set rather arbitrarily in the meV range and their anisotropy can be then tuned to match the observed zero-field FMR energy gaps and the measured field dependence of the frequencies of the FMR modes. Although the obtained fit parameters might not be unique, the modeling can answer the important conceptual question on whether the scenario of interacting FM-AFM lattices can at all explain the occurrence of the additional FMR branch in the excitation spectrum of \LCS.         

\begin{figure}
	\centering
	\includegraphics[width=\linewidth]{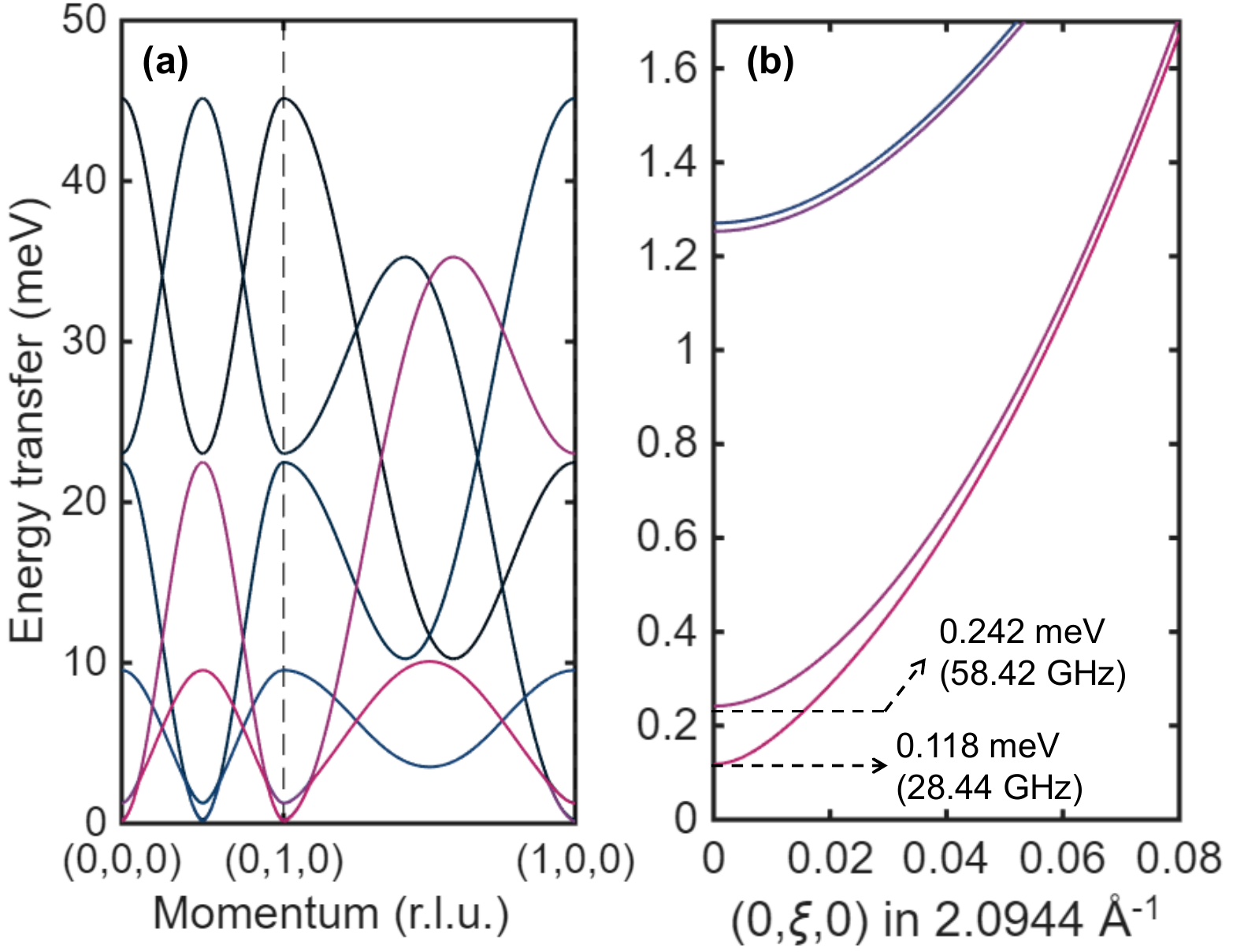}
	\caption{Modeled spin wave excitation spectrum with parameters from Table~\ref{tab:model_param}. (a) General view. (b) Low-energy modes near the zone center (0 0 0). Magnon excitation energies at $\mathbf{q} = 0$ probed by FMR are indicated by dashed arrows.}
	\label{fig:spin_waves}
\end{figure}

The resulting spin wave excitation spectrum in this model, whose magnetic unit cell comprises 16 atoms, is rather complex and consists of several magnon modes dispersing in the magnetic Brillouin zone (Fig.~\ref{fig:spin_waves}(a)). The low-energy part of these excitations near the zone center is shown in Fig.~\ref{fig:spin_waves}(b). Importantly, there are two non-degenerate $\mathbf{q} = 0$ excitations at lowest energies which are relevant for FMR. As the modeling shows, the non-degeneracy of these magnon modes is essentially due to the finite  FM--AFM interlattice coupling (Eq.~(\ref{eq:J12matrix})). In its absence they would have the same energy corresponding to the FMR frequency of a one sublattice ferromagnet. The field dependence of these two modes for $\mathbf{H}\parallel \mathbf{c}$ (in-plane hard axis) is shown in Fig.~\ref{fig:model_branches}. The best possible agreement with the experimental data was achieved for the set of parameters listed in Table~\ref{tab:model_param}.

\setlength{\tabcolsep}{12pt} 
\renewcommand{\arraystretch}{1.5} 
\begin{table}[]
	\caption{Exchange parameters according to Eqs.~(\ref{eq:J1matrix})-(\ref{eq:J12matrix}) set to model the FMR branches in Fig.~\ref{fig:model_branches} (in meV). }
	\label{tab:model_param}
		\begin{tabular}{|c|c|c||c|c|}
			\hline
			$J^{xx}_{1b}$ & $J^{yy}_{1b}$ & $J^{zz}_{1b}$ & $J_{2b}$      & $J_{2c}$      \\ \hline
			-7.0484       & -6.7          & -6.7          & -6.7          & 6.7     \\ \hline\hline
			$J^{xx}_{1c}$ & $J^{yy}_{1c}$ & $J^{zz}_{1c}$ & $J^{xy}_{12}$ & $J^{yx}_{12}$ \\ \hline
			-6.7          & -7.035        & -6.7          & $0.14$    & $ 0.14$    \\ \hline
		\end{tabular}
\end{table}

Given a rather approximate character of the model, the overall agreement between the calculated and experimentally determined FMR branches of \LCS is reasonably good. The only noticeable deviations appear at small fields (Fig.~\ref{fig:model_branches}, inset). In particular, the calculated low-frequency mode exhibits a sharp softening  with increasing the field strength at the field $\sim 0.2$\,T indicating reorientation of the spins in the FM sublattice from the easy $b$-axis direction towards the hard $c$-axis. Experimentally, this mode cannot be resolved below this critical field, but it emerges practically at this field with the frequency corresponding to the zero-field excitation energy 0.12\,meV (29\,GHz) of the lowest in energy magnon branch at $\mathbf{q} = 0 $ in Fig.~\ref{fig:spin_waves}(b).  The modeled high-frequency FMR branch also features a small anomaly at this field. However, it cannot be resolved experimentally, possibly due to the broadness of the corresponding FMR mode. Nevertheless, it still can be traced down to almost zero field where it has the frequency corresponding to the excitation energy 0.24\,meV (58\,GHz) of  the second magnon branch in Fig.~\ref{fig:spin_waves}(b).                                    

As discussed above, the lack of knowledge of the values of the exchange couplings in \LCS leads to some ambiguity in the choice of parameters. In particular, it prevents a reliable prediction of the dispersion of the spin wave excitations in the $q$-space. The example shown in Fig.~\ref{fig:spin_waves} presents just one of many possibilities. However, the analysis of the model shows that the magnon excitation energies at the Brillouin zone center probed by FMR spectroscopy depend only on the deviation of exchange interactions in the FMR lattice from the isotropic limit.
Thus, exchange anisotropy can be unambiguously determined by measuring the field dependence of the FMR modes. In the case of \LCS the anisotropy turns out to be relatively small, of the order of 5\,\% (Table~\ref{tab:model_param}), which explains the smallness of the zero field excitation energy gaps. Most importantly, the model successfully reproduces the two experimentally observed FMR branches with distinct energies that are not expected for a one sublattice ferromagnet. The energy separation between the branches in this model is essentially due to the finite coupling between the FM and orthogonal AFM sublattices and can be fine tuned to match the experiment (Fig.~\ref{fig:model_branches}). Therefore, the above analysis of the experimental data in the frame of this model strongly supports on the phenomenological level the scenario of coexisting orthogonal FM and AFM sublattices in \LCS proposed in Ref.~\cite{Granado107204} and underscores unconventional magnetism of this compound that obviously deserves further theoretical and experimental studies.

\section{Conclusion}
In summary, we have performed a detailed ESR/FMR spectroscopic study of poly- and single-crystalline samples of the metallic ferromagnet \LCS with the layered crystallographic structure over a broad range of excitation frequencies and temperatures. The samples were carefully characterized by the x-ray diffraction and static magnetization measurements. The structural data reveals a pronounced anomaly in the temperature dependence of the lattice constants at the ferromagnetic ordering temperature $T_{\rm C} \simeq 126$\,K, suggesting a strong magneto-elastic coupling in this material. ESR measurements above $T_{\rm C}$ reveal the presence of static on the ESR timescale FM correlations persistent up to $\sim 200$\,K. This result indicates a quasi-2D character of the spin system in \LCS which can be expected given the layered type of the crystallographic structure of this compound.  Below $T_{\rm C}$, temperature and frequency dependent FMR measurements have enabled us to identify the FMR excitation modes and to establish their magnetic field dependence (FMR branches). 

A remarkable result of this study is the observation at low temperatures of a second branch in the FMR excitation spectrum, which is not expected for a one sublattice ferromagnet which \LCS is supposed to be. It suggests the presence of at least two magnetic sublattices in this compound. This conjecture was put forward already before to explain a significant alternating tilting of the FM ordered moments by assuming the presence of orthogonal FM and AFM sublattices built of localized and itinerant Cr $d$ electrons, respectively \cite{Granado107204}. To verify this concept, we have constructed a minimal phenomenological model of interating FM-AFM sublattices and numerically calculated its spin wave excitation spectrum. We have shown that the model naturally yields two, distinct in energy FMR branches for a given magnetic field direction, and that it is possible to achieve even a good quantitative agreement between the modeled and experimentally observed FMR branches, albeit the choice of the model parameters might not be unique. Our results uncover unconventional aspects of magnetism in the metallic ferromagnet \LCS and call for further theoretical and experimental studies to elucidate the microscopic origin of its canted magnetic structure and FMR excitation spectrum.

\begin{acknowledgments}
The authors gratefully acknowledge financial support by the Science and Engineering Research Board (SERB), India (Grant No.~CRG/2022/000997), by the Deutsche Forschungsgemeinschaft (DFG) through Project No. 499461434 (AL 1771/8-1), and within the Collaborative Research Center SFB 1143 ``Correlated Magnetism - From Frustration to Topology'' (project-id 247310070), and the Dresden-W\"urzburg Cluster of Excellence (EXC 2147) ``ctd.qmat - Complexity and Topology in Quantum Matter'' (project-id 390858490). SM, RK, VS, and RN would like to acknowledge Science and Engineering Research Board (SERB), India bearing Grant No. CRG/2022/000997 for financial support.

\end{acknowledgments}

\section{Appendix}

\setcounter{table}{0}
\renewcommand{\thetable}{A\arabic{table}}
\setcounter{figure}{0}
\renewcommand{\thefigure}{A\arabic{figure}}
\setcounter{equation}{0}
\renewcommand{\theequation}{A\arabic{equation}}

\subsection{Analysis of the ESR spectra}
\label{section:analysis_spectra}

\subsubsection{X-band ESR}

In the X-band ESR setup, the field derivative of the ESR absorption signal is recorded. The experimental spectra, characterized by a single resonance line, were fitted with the first derivative of the Lorentzian function:
\begin{align}
	S_\mathrm{abs} =& \frac{-2x}{(1+x^2)^2} + \frac{-2y}{(1+y^2)^2},
\end{align}
where
\begin{align}
	x = \frac{2(H+H_{res})}{\Delta H}, \hspace{5 pt}
	y = \frac{2(H-H_{res})}{\Delta H}\ .
	\label{x_y}
\end{align}

The contribution involving $y$ is necessary for precisely fitting broad spectra. In the ESR setup the linearly polarized microwaves are used which can be decomposed into the right and left circularly polarized components. These two components produce ESR signals in the positive and negative magnetic fields, respectively, which may overlap in the case of broad lines. The $y$ term accounts for the contribution to the total measured signal of the component from the negative field. 

\subsubsection{HF-ESR}

In the HF-ESR setup, the mixing of absorption $S_\mathrm{abs}$ and dispersion $S_\mathrm{disp}$ components of the detected signal is unavoidable due to the complex impedance of the broadband probehead. Generally, the HF-ESR signal with intermixed $S_\mathrm{abs}$ and $S_\mathrm{disp}$ components and a Lorentzian line profile can thus be represented as:
\begin{align}
	S_\mathrm{D}^\mathrm{amp}=\frac{2A_0}{\pi}(S_\mathrm{abs}\sin\alpha+S_\mathrm{disp}\cos\alpha)+S_o+S_s\cdot H 
	\label{eq:lorentzian}
\end{align}
where $A_0$ is the amplitude of the signal, $S_0$ is the offset, and $S_s$ is the linear background in the detected signal. Here $S_\mathrm{abs}$ and $S_\mathrm{disp}$ are related by the Kramers-Kronig relation and can be expressed as:
\begin{align}
	S_\mathrm{abs} =& \frac{1}{1+x^2} + \frac{1}{1+y^2}\\
	S_\mathrm{disp} =& \frac{x}{1+x^2} + \frac{y}{1+y^2}\ ,
\end{align}
where $x$ and $y$ are defined in  Eq.~(\ref{x_y}). The pure absorption component of the as-measured HF-ESR spectra was rectified by varying the mixing parameter $\alpha$.

\onecolumngrid
\subsection{Additional magnetization and FMR data}
\phantomsection
\label{section:additional_data}

\setcounter{table}{0}
\renewcommand{\thetable}{B\arabic{table}}
\setcounter{figure}{0}
\renewcommand{\thefigure}{B\arabic{figure}}
\setcounter{equation}{0}
\renewcommand{\theequation}{B\arabic{equation}}

\begin{figure*}[!h]
	\centering
	\includegraphics[width=\linewidth]{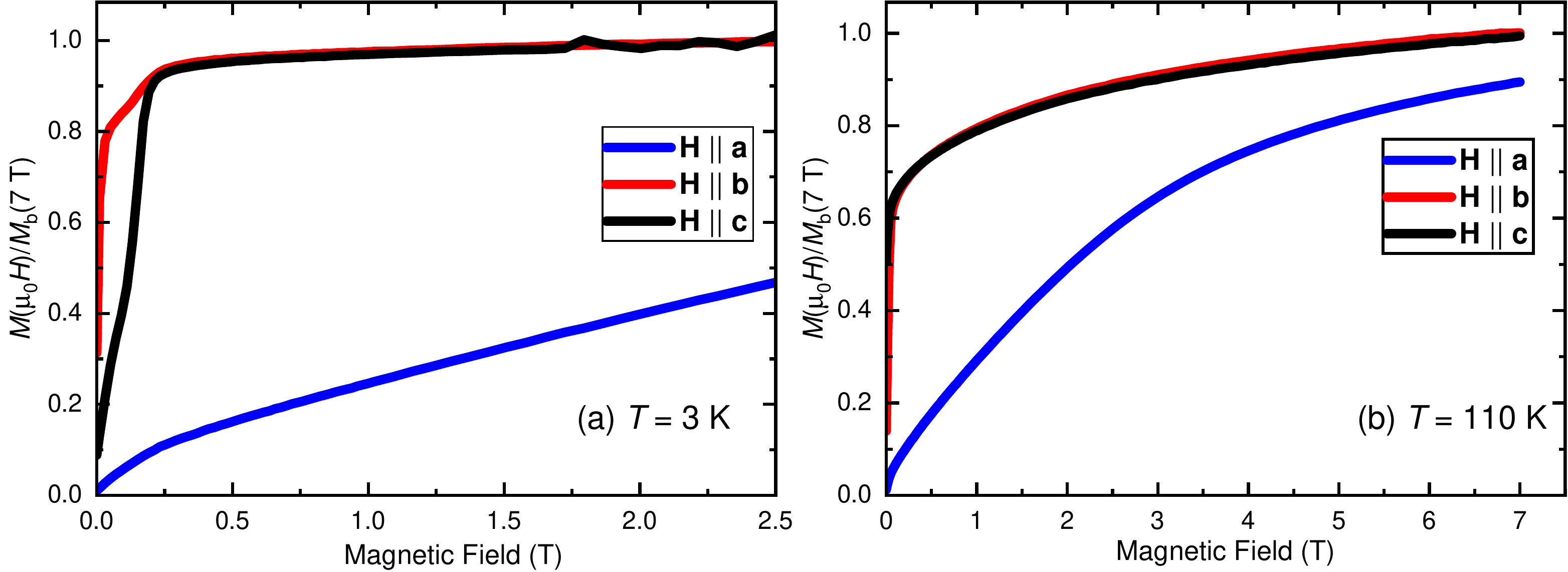}
	\caption{Field dependence of isothermal magnetization normalized to the value of $M_{\rm b}$ at $\mu_{0}H = 7$\,T for (a) 3\,K and (b) 110\,K along the \HIIa, \HIIb and \HIIc configurations. 
	These measurements were performed on the single crystal of \LCS used for ESR studies in order to identify the crystallographic axes. A nonmonotonic behavior of $M(H)$ at $ T = 3$\,K in small fields applied in the $bc$-plane indicates a metamagnetic transition due to an additional easy-axis anisotropy in this plane.  	
}
	\phantomsection
	\label{fig:MH}
\end{figure*}

\begin{figure*}
	\includegraphics[width=\linewidth]{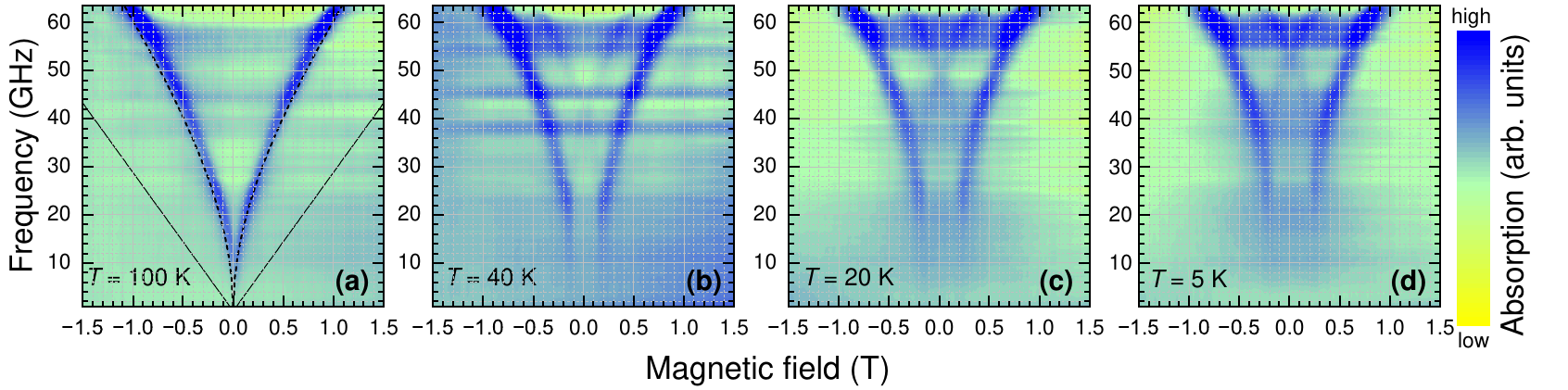}
	\caption{Two-dimensional color-coded view of the frequency swept FMR spectra of a single crystal of \LCS for $\mathbf{H}\parallel \mathbf{b}$ field geometry at (a) 100\,K, (b) 40\,K, (c) 20\,K, and (d) 5\,K. Dashed curve in (a) is the fit of the data to Eq.~(\ref{eq:LWST1}), and the thin solid line represents the paramagnetic branch according to Eq.~(\ref{eq:PM}). Horizontal stripes are artefacts caused by parasitic resonances in the coplanar waveguide. }
	\phantomsection
	\label{fig:coplanarBaxis}
\end{figure*}

\FloatBarrier
\twocolumngrid

\bibliography{ref_LaCrSb3}

\end{document}